Original Paper

# Challenges in Using mHealth Data From Smartphones and Wearable Devices to Predict Depression Symptom Severity: Retrospective Analysis

Shaoxiong Sun[1], PhD; Amos A Folarin[1,2,3,4,5], PhD; Yuezhou Zhang[1], PhD; Nicholas Cummins[1], PhD; Rafael Garcia-Dias[6], PhD; Callum Stewart[1], MSc; Yatharth Ranjan[1], MSc; Zulqarnain Rashid[1], PhD; Pauline Conde[1], BSc; Petroula Laiou[1], PhD; Heet Sankesara[1], BSc; Faith Matcham[7,8], PhD; Daniel Leightley[7], PhD; Katie M White[7], BSc; Carolin Oetzmann[7], MSc; Alina Ivan[7], MSc; Femke Lamers[9,10], PhD; Sara Siddi[11,12,13], PhD; Sara Simblett[14], PhD; Raluca Nica[15,16], MD; Aki Rintala[17,18], PhD; David C Mohr[19], PhD; Inez Myin-Germeys[17], PhD; Til Wykes[14,20], PhD; Josep Maria Haro[11,12,13], MD; Brenda W J H Penninx[9,10], PhD; Srinivasan Vairavan[21], PhD; Vaibhav A Narayan[22], PhD; Peter Annas[23], PhD; Matthew Hotopf[3,7,20], PhD; Richard J B Dobson[1,2,3,4,5], PhD; RADAR-CNS Consortium[24]

[1]Department of Biostatistics & Health Informatics, Institute of Psychiatry, Psychology and Neuroscience, King's College London, London, United Kingdom

[2]Institute of Health Informatics, University College London, London, United Kingdom

[3]NIHR Biomedical Research Centre at South London and Maudsley, NHS Foundation Trust, London, United Kingdom

[4]Health Data Research UK London, University College London, London, United Kingdom

[5]NIHR Biomedical Research Centre at University College London Hospitals, NHS Foundation Trust, London, United Kingdom

[6]Department of Psychosis Studies, Institute of Psychiatry, Psychology and Neuroscience, King's College London, London, United Kingdom

[7]Department of Psychological Medicine, Institute of Psychiatry, Psychology and Neuroscience, King's College London, London, United Kingdom

[8]School of Psychology, University of Sussex, Falmer, United Kingdom

[9]Department of Psychiatry, Amsterdam UMC, Vrije Universiteit, Amsterdam, Netherlands

[10]Mental Health Program, Amsterdam Public Health Research Institute, Amsterdam, Netherlands

[11]Teaching Research and Innovation Unit, Parc Sanitari Sant Joan de Déu, Fundació Sant Joan de Déu, Barcelona, Spain

[12]Centro de Investigación Biomédica en Red de Salud Mental, Madrid, Spain

[13]Faculty of Medicine and Health Sciences, Universitat de Barcelona, Barcelona, Spain

[14]Department of Psychology, Institute of Psychiatry, Psychology and Neuroscience, King's College London, London, United Kingdom

[15]RADAR-CNS Patient Advisory Board, King's College London, London, United Kingdom

[16]The Romanian League for Mental Health, Bucharest, Romania

[17]Department of Neurosciences, Center for Contextual Psychiatry, Katholieke Universiteit Leuven, Leuven, Belgium

[18]Physical Activity and Functional Capacity Research Group, Faculty of Health Care and Social Services, LAB University of Applied Sciences, Lahti, Finland

[19]Center for Behavioral Intervention Technologies, Department of Preventive Medicine, Northwestern University, Chicago, IL, United States

[20]South London and Maudsley NHS Foundation Trust, London, United Kingdom

[21]Janssen Research and Development LLC, Titusville, NJ, United States

[22]Davos Alzheimer's Collaborative, Wayne, PA, United States

[23]H Lundbeck A/S, Copenhagen, Denmark

[24]See Acknowledgments

**Corresponding Author:**
Shaoxiong Sun, PhD
Department of Biostatistics & Health Informatics, Institute of Psychiatry, Psychology and Neuroscience
King's College London
Social Genetic and Developmental Psychiatry Centre
Memory Lane
London, SE5 8AF
United Kingdom
Phone: 44 02078480951
Email: shaoxiong.sun@kcl.ac.uk

## Abstract

**Background:** Major depressive disorder (MDD) affects millions of people worldwide, but timely treatment is not often received owing in part to inaccurate subjective recall and variability in the symptom course. Objective and frequent MDD monitoring can improve subjective recall and help to guide treatment selection. Attempts have been made, with varying degrees of success, to explore the relationship between the measures of depression and passive digital phenotypes (features) extracted from smartphones and wearables devices to remotely and continuously monitor changes in symptomatology. However, a number of challenges exist for the analysis of these data. These include maintaining participant engagement over extended time periods and therefore understanding what constitutes an acceptable threshold of missing data; distinguishing between the cross-sectional and longitudinal relationships for different features to determine their utility in tracking within-individual longitudinal variation or screening individuals at high risk; and understanding the heterogeneity with which depression manifests itself in behavioral patterns quantified by the passive features.

**Objective:** We aimed to address these 3 challenges to inform future work in stratified analyses.

**Methods:** Using smartphone and wearable data collected from 479 participants with MDD, we extracted 21 features capturing mobility, sleep, and smartphone use. We investigated the impact of the number of days of available data on feature quality using the intraclass correlation coefficient and Bland-Altman analysis. We then examined the nature of the correlation between the 8-item Patient Health Questionnaire (PHQ-8) depression scale (measured every 14 days) and the features using the individual-mean correlation, repeated measures correlation, and linear mixed effects model. Furthermore, we stratified the participants based on their behavioral difference, quantified by the features, between periods of high (depression) and low (no depression) PHQ-8 scores using the Gaussian mixture model.

**Results:** We demonstrated that at least 8 (range 2-12) days were needed for reliable calculation of most of the features in the 14-day time window. We observed that features such as sleep onset time correlated better with PHQ-8 scores cross-sectionally than longitudinally, whereas features such as wakefulness after sleep onset correlated well with PHQ-8 longitudinally but worse cross-sectionally. Finally, we found that participants could be separated into 3 distinct clusters according to their behavioral difference between periods of depression and periods of no depression.

**Conclusions:** This work contributes to our understanding of how these mobile health–derived features are associated with depression symptom severity to inform future work in stratified analyses.

*(J Med Internet Res 2023;25:e45233)* doi: [10.2196/45233](10.2196/45233)

**KEYWORDS**

mobile health; depression; digital phenotypes; behavioral patterns; missing data; smartphones; wearable devices; mobile phone

## Introduction

### Background

Major depressive disorder (MDD) is a mental disorder with an estimated lifetime risk worldwide of 16.6% to 17.1% [1,2]. It leads to high medical costs, morbidity, distress, and mortality [3-5]. Approximately 60% of the people meeting the diagnostic criteria of MDD experience severely impaired function and considerably degraded self-care and life quality [6]. Unsurprisingly, given these factors, MDD is the second-leading cause of chronic disease burden as measured by years lived with disability [7].

Two main treatments, psychotherapy and pharmacotherapy, are available for the management of MDD with varying degrees of success [7]. However, according to a study, >50% of the people with MDD do not receive timely treatment [8]. One reason is that the current assessment of symptom severity can be compromised by recollection errors when patients are asked to fill in subjective and retrospective self-reported questionnaires [9]. Another main reason is that the symptom course is often not known to clinicians, and variability in the symptom course adds noise to the study of treatment [10]. A more objective and frequent monitoring of mental health status may not only improve subjective recall but also help to guide treatment selection.

Attempts have been made to leverage smartphones, which are now ubiquitous in mental health status monitoring [11-14]. Sensors on these devices have been used to capture behaviors such as mobility and smartphone use [15]. Specifically, digital phenotypes (features) derived from GPS mobility and smartphone use were extracted and correlated with the level of depression assessed by the 8-item Patient Health Questionnaire (PHQ-8) depression scale [12,16,17]. Wearable devices such as Fitbit have also been used to calculate mobility and sleep features [18-20]. Small to moderate correlations between PHQ-8 scores and the features were found in cross-sectional studies where participants were monitored for a relatively short time period with a limited number of PHQ-8 measures per participant [11,12]. Discrepancies in findings could be due to study design, population, sample size, and data [14].

A number of challenges exist for these studies. First, participant engagement is typically a challenge for mobile health (mHealth) studies, which presents a missing data problem for analysis [21]. Therefore, it is important to understand what constitutes an acceptable level of missing data when summarizing data collected in a time window. A second challenge lies in differentiating between cross-sectional and longitudinal correlations of the features and depression symptom severity. Although cross-sectional and longitudinal correlations have been reported in different studies [12,18,22], it is still unclear whether a feature is more suited to detect within-individual longitudinal changes or screen individuals at higher risk. A third challenge exists in understanding the heterogeneity with which depression manifests itself in terms of behavioral patterns. It is very common that people with MDD only exhibit an individualized subset of the whole spectrum of MDD symptoms [7,23]. Furthermore, several symptoms of depression can be experienced in opposite extremes; for example, sleep problems can be experienced in the form of hypersomnia or insomnia [6,24]. The difference in the exhibited behavioral patterns complicates the correlation analysis assessing the associations between the features and symptom severity.

### Objectives

In this work, we explored these 3 challenges. We first investigated the impact of data missingness on the quality of a few representative features capturing mobility, sleep, and smartphone use. Next, we examined both cross-sectional and longitudinal correlations between the features and PHQ-8 scores to better understand their utility for disease screening or tracking disease progression. Finally, we sought to understand the heterogeneity in depression by clustering participants into distinctive groups based on behavioral patterns quantified by the mHealth-derived features.

## Methods

### Data Collection

This work is part of the Innovative Medicines Initiative (IMI)-2 Remote Assessment of Disease and Relapse–Central Nervous System (RADAR-CNS) major program [25], which aimed to evaluate remote monitoring using wearables and smartphones in 3 central nervous system disorders (MDD, epilepsy, and multiple sclerosis) [26-31]. The Remote Assessment of Disease and Relapse–Major Depressive Disorder (RADAR-MDD) study [26] recruited 623 participants with a recent history of recurrent MDD at 3 sites: King's College London in London, United Kingdom; Centro de Investigación Biomédica en Red (CIBER) in Barcelona, Spain; and Vrije Universiteit Medisch Centrum (VUMC) in Amsterdam, the Netherlands. In this analysis, of the 623 participants with MDD, we included 479 (76.9%) recruited between November 2017 and January 2020 at the 3 clinical sites to exclude the impact of COVID-19 [19,32]. Their demographic information is presented in Table 1. We also provide the demographics for the whole sample in the RADAR-MDD study as a comparison. We performed chi-square tests for the distribution of categories in terms of gender, employment status, marital status, and medication status. We also performed Wilcoxon rank sum tests on age and length of education. A $P$ value of <.05 was deemed statistically significant. No significant difference was observed (gender: $P$=.58; employment status: $P$=.65; marital status: $P$=.89; medication for depression: $P$=.99; age: $P$=.69; and length of education: $P$=.44), which indicates that the cohort used in this analysis is representative of the full cohort.

Passive participant data, that is, data that did not require active participant engagement (eg, GPS, step count, and sleep data), were collected on a 24/7 basis through a smartphone and a Fitbit device. We used the participants' own Android smartphones where available; otherwise, we provided participants with a Motorola G5, G6, or G7 smartphone. Wrist-worn devices, either Fitbit Charge 2 or Fitbit Charge 3, were given to participants, and they were asked to wear the device on their nondominant hand. Active participant data, that is, data obtained by administering fortnightly app-delivered questionnaires (eg, the PHQ-8) to participants, were also collected. The first PHQ-8 score was collected at baseline (day 1). We did not include the baseline PHQ-8 scores in the analysis. The data collection and management were handled by the open-source mHealth platform Remote Assessment of Disease and Relapse–base (RADAR-base) [27].

In this analysis, the participants had a median of 312 (IQR 126-453) days of stay in the study, completing a median of 9 (IQR 4-17) PHQ-8 questionnaires. We used the PHQ-8 scores to measure depression symptom severity. The PHQ-8 has been shown to be a useful depression measure for population-based studies [33], and it is also used in mHealth studies [18,20,34].

**Table 1.** Participant demographics.

| Demographics | Cohort before COVID-19 pandemic[a] (n=479) | Full cohort (N=623) |
| --- | --- | --- |
| Age (years), median (IQR) | 50.0 (33.0-59.0) | 49.0 (33.0-59.0) |
| **Gender, n (%)** | | |
|     Man | 122 (25.5) | 152 (24.4) |
|     Woman | 357 (74.5) | 471 (75.6) |
| **Employment status, n (%)** | | |
|     Employed | 231 (48.2) | 305 (49) |
|     Unemployed | 64 (13.4) | 82 (13.1) |
|     Student | 45 (9.4) | 67 (10.8) |
|     Retired | 103 (21.5) | 119 (19.1) |
|     Other | 36 (7.5) | 50 (8) |
| Length of education (years), median (IQR) | 15.0 (12.0-19.0) | 16.0 (13.0-19.0) |
| **Marital status, n (%)** | | |
|     Single | 168 (35.1) | 223 (35.8) |
|     Married or cohabiting | 229 (47.8) | 291 (46.7) |
|     Divorced, separated, or widowed | 82 (17.1) | 109 (17.5) |
| **Medication for depression, n (%)** | | |
|     Yes | 315 (65.8) | 408 (65.5) |
|     No | 77 (16.1) | 100 (16.1) |
|     Not reported | 87 (18.1) | 115 (18.5) |

[a]Until January 31, 2020.

## Ethics Approval

The RADAR-MDD study was conducted in accordance with the Declaration of Helsinki and Good Clinical Practice guidelines, adhering to principles outlined in the National Health Service (NHS) Research Governance Framework for Health and Social Care (second edition). Ethics approval was obtained in the United Kingdom from the Camberwell St Giles Research Ethics Committee (17/LO/1154), in Spain from the CEIC Fundació Sant Joan de Deu (CI PIC-128-17), and in the Netherlands from the Medisch Ethische Toetsingscommissie VUmc (METc VUmc registratienummer 2018.012–NL63557.029.17). All participants signed informed consent.

## Feature Extraction

Existing research shows that reduced mobility is associated with depression [12,16,17]. In this work, we extracted homestay duration and maximum traveled distance from home from the smartphone GPS (relative location) data, as implemented in the study by Sun et al [19]. The smartphone-derived location data had a sampling period of 5 minutes by default, with longer sampling durations dependent on network connectivity and battery level. Spurious location coordinates were identified and removed if they differed from the preceding and following coordinates by >5°. Home location was determined daily by clustering location data between 8 PM and 4 AM using the mean coordinate of the cluster containing the last coordinate. This choice was made because the largest cluster may not be the home location for a single night, but we assumed that the last location before the smartphones shut down had a higher probability to be the home location for that night. The clustering was implemented using density-based spatial clustering of applications with noise [35]. A duration gated by 2 adjacent coordinates was regarded as a valid homestay duration on the condition that both coordinates were no further than 200 meters from the home location. One-time durations of >1 hour were excluded owing to the large proportion of missing data when compared with the 5-minute sampling period. All valid homestay durations between 8 AM and 11 PM were summed to calculate daily homestay duration. Daily maximum traveled distance from home was also computed based on the coordinates in the same time period to quantify mobility. In addition, we computed the daily step count as the total steps a participant walked every day.

Features from the Fitbit sleep data have been shown to be correlated with depression symptom severity [20]. In this work, sleep episodes were determined using Fitbit-detected continuous sleep stages (light, deep, and rapid eye movement [REM]), which were sampled every 30 seconds by default when sleep was detected. The longest sleep episode was used when ≥2 sleep episodes were found. Sleep duration was computed as the length of the sleep episode. Time in bed was then calculated as the sum of awake duration and sleep duration, and sleep efficiency was calculated as the ratio between sleep duration and time in bed. Sleep onset time and sleep offset time were determined using the start of the first sleep stage and the end of the sleep stage, respectively. Wakefulness after sleep onset (WASO) and wakefulness after sleep offset (WASF) were calculated as the awake durations before the first sleep stage and after the last sleep stage, respectively.

To quantify smartphone use, previously shown to be related to depression [12,15], we extracted features from smartphone user interaction and use event data streams as implemented in the study by Sun et al [19]. From smartphone user interaction data, we calculated the daily times and daily total duration of the smartphone being in the unlock status. Specifically, smartphone unlock duration sum (daily total duration) was calculated by summing time periods starting with the unlocked state and ending with the standby state. All one-time (continuous) unlock intervals that were >4 hours in duration were excluded because they might be the result of a missing standby state, unintentionally leaving the phone unlocked, or watching videos. Furthermore, we computed the minimum and maximum of all valid one-time unlock intervals and the median of all time gaps between 2 adjacent unlocking activities. To extract features from smartphone use event data, apps were first grouped into classes according to the classification listed on the Google Play Store [36]. We further combined similar classes into 3 larger categories: social apps, attention apps, and game apps. Social apps included the classes of "Social," "Communication," and "Dating"; attention apps, which require users to focus on the app for extended time, included the classes of "Productivity," "Books & Reference," "Education," "News & Magazines," "Business," and "Art & Design"; and game apps included the classes of "Action," "Action & Adventure," "Adventure," "Arcade," "Card," "Casino," "Brain games," "Board," "Puzzle," "Simulation," "Racing," "Role playing," "Strategy," "Pretend play," "Word," "Casual," "Educational," "Trivia," and "Sports." The daily times and durations of using the social, attention, and game apps were computed.

A full list of the extracted features is presented in Table 2. These features were extracted for each participant daily. The daily features were calculated using the data from 6 AM on the day to 6 AM the next day for all features except for sleep features, where the longest nonstop sleep episode was selected. When no data were found in a data stream for a participant on a day, we recorded the features derived from that data stream on that day as missing. Note that this work did not enumerate and implement all existing features in the literature. The focus of this study was to provide a methodology and exemplify it using a few representative features.

**Table 2.** Full list of the extracted features.

| Categories, data streams, and features | Extraction |
| --- | --- |
| **Mobility** | |
|   **Smartphone GPS** | |
|     Homestay duration | The duration of staying at home |
|     Maximum traveled distance from home | The maximum traveled distance from home location |
|   **Fitbit step count** | |
|     Step count | Daily total of Fitbit step counts |
| **Sleep** | |
|   **Fitbit sleep** | |
|     Sleep duration | Daily total duration of Fitbit-detected sleep stages |
|     Time in bed | Daily total duration of staying in bed |
|     Sleep efficiency | The ratio between sleep duration and time in bed |
|     WASO[a] | Awake duration after sleep onset and before sleep offset |
|     WASF[b] | Awake duration after sleep offset |
|     Sleep onset time | The time of falling asleep |
|     Sleep offset time | The time of waking up |
| **Smartphone use** | |
|   **Smartphone user interaction** | |
|     Unlock times and duration sum | The daily total times and duration of the smartphone spent in the unlocked state |
|     Unlock duration min and max | The one-time minimum and maximum duration of the smartphone spent in the unlocked state |
|     Median interval between 2 unlocks | The median of all time gaps between 2 adjacent unlocking activities |
|   **Smartphone app use event** | |
|     Social app use times and duration | The total times and duration of the smartphone spent on social apps |
|     Attention app use times and duration | The total times and duration of the smartphone spent on attention-requiring apps |
|     Game app use times and duration | The total times and duration of the smartphone spent on game apps |

[a]WASO: wakefulness after sleep onset.

[b]WASF: wakefulness after sleep offset.

## The Impact of Missing Data on Feature Quality

We investigated the impact of the number of days included when calculating summary statistics (mean and median) for 14-day windows before the completion of the PHQ-8. We chose the 14-day window based on the fact that the PHQ-8 asks about one's depression symptoms in the past 14 days, which is also in line with other existing works [37,38]. We considered the 14-day periods with no missing daily data and simulated hypothetical data missingness by selecting all subsets of $i$ days' data from the 14-day period, where $i$=1, 2,..., 13. We used 2-way random effects intraclass correlation coefficient (ICC) [39,40] to assess the absolute agreement between the $i$ days and the complete data for the 14-day period. A higher ICC among all combinations of $i$-day summary statistics indicates a higher absolute agreement and reliability. We used 0.9 as the threshold to determine the minimal number of days [41]. We further assessed the agreement in the summary statistics calculated between $i$-day data and 14-day data. This was done by using the Bland-Altman analysis, which is often used to quantify the agreement between 2 methods of measurement [42-44]. The mean of their difference, the SD of their difference (precision), and the precision ratio were calculated to indicate variability. The precision ratio was computed as the ratio between precision and the mean feature values given $i$-day data. A larger precision ratio indicates a larger variability.

## Cross-Sectional and Longitudinal Correlations of the Extracted Features With Depression Symptoms

We examined both the cross-sectional and longitudinal correlations between the PHQ-8 scores and extracted features. For the cross-sectional correlation, we first averaged the features for each participant over all valid periods where at least 8 days of data were present. This choice is justified in the *Results* section. We also averaged PHQ-8 scores over these periods. This led to 1 data pair per participant holding the assumption of independence of observation [45]. We then calculated Spearman correlation coefficients on these data pairs. We chose Spearman correlation to assess the monotonic relationships (either linear or nonlinear) between them [46]. For the longitudinal correlation, we used repeated measures correlation and linear mixed effects models, where multiple PHQ-8 scores per participant were taken as outcome variables. Repeated measures correlation is a statistical technique for determining the common within-individual association for repeated and paired measures for multiple participants [47]. It assumes common within-individual variance and estimates the commonly shared regression slope between the paired measures, while not violating the assumption of independence of observation [47]. Linear mixed effects models differ from repeated measures correlations in that they simultaneously model the variance from different (fixed and random) sources [48]. In addition to the coefficient and 2-tailed $t$ value from these methods, respectively, we also presented rankings based on the coefficient and $t$ value to better understand the relative relevance of the features. We chose to use the $t$ value in the linear mixed effects model for its capability to indicate the strength of the relationship between the features and PHQ-8 scores and for its invariance to the scale of the features [28].

**Clustering of Participants Based on Behavioral Changes During Periods of Depression**

To begin with, we quantified the behavioral changes between periods of high depression scores and periods of low depression scores. To do so, we only included participants who had at least 1 period of depression and 1 period of no depression as dictated by PHQ-8 scores with a threshold score of 10 (PHQ-8 score ≥10 and PHQ-8 score <10, respectively). We required the presence of at least 8 days of passive data out of the 14 days before a PHQ-8 assessment. This choice is justified in the *Results* section. We then put together data from all respective periods for each of the features and for each of the states (depression and no depression). By doing so, we were able to investigate the behavioral patterns in the depression or no-depression states aggregated from the corresponding periods. To reliably assess the difference in the distribution of the behavioral features in these 2 states, we used an indicator analogous to the effect size of the rank sum test. The indicator was computed as the test statistic in the rank sum test divided by the square root of the total number of days in both depression and no-depression states. This processing led to 1 data point per feature per participant.

On these features in the form of the processed difference, we applied a feature selection approach to identify the most informative features for clustering. The approach we used was the principal feature analysis (PFA), which is based on the principal component analysis (PCA) [49]. In the PFA, the number of principal components ($q$) was first determined based on retained variability from the PCA. In this work, we used the explained variance ratio of 80% from the PCA as the requirement of retained variability. This was to strike a balance between keeping useful information and rejecting noise. Next, $p$ eigenvectors corresponding to the $p$ largest eigenvalues from the PCA were clustered using the k-means algorithm. For each cluster, the feature with the eigenvector closest to the center of the cluster was selected. Here, we selected $p = q + 2$ in line with the study by Lu et al [49]. The reason for choosing $p$ larger than $q$ is that a slightly larger number of features is needed for retaining the same variability as in the PCA [49]. These selected features were taken forward for the analysis of the heterogeneity.

To discover the underlying heterogeneity of behaviors and symptoms, we used a clustering algorithm on the processed difference of behavioral features derived from each participant. Specifically, after standardizing the features, we applied a Gaussian mixture model. A Gaussian mixture model assumes that all data are generated from a mixture of a finite number of Gaussian distributions, which can be regarded as clusters [50]. The optimal number of clusters was determined using the silhouette coefficient, a measure of cluster cohesion and separation [51]. A higher silhouette coefficient indicates a better cluster cohesion and separation. Here, we reported the center and weight of each cluster. Parallel plots on the standardized data were provided to further visualize the different clusters. This standardization was carried out for each feature in all clusters put together. To further study the group difference, we also calculated the repeated measures correlation coefficients within each cluster.

This work was implemented in Python (version 3.7.4; Python Software Foundation). For feature extraction, key packages included *pandas*, *NumPy*, and *scikit-learn*. For the impact of missing data, key packages included *pandas*, *NumPy*, and *Pingouin*. For the nature of the correlation, key packages included *pandas*, *NumPy*, and *Pingouin*, and *statsmodels*. For the clustering of the participants based on behavioral patterns, key packages included *pandas*, *NumPy*, and *scikit-learn*.

## Results

### The Impact of Missing Data on Feature Quality

For smartphone GPS, median data availability was 13 (IQR 1-14) days out of the maximum of 14 days. For Fitbit step count, median data availability was 14 (IQR 0-14) days. For smartphone user interaction and app use, median data availability was 13 (IQR 0-14) days. For Fitbit sleep, median data availability was 11 (IQR 3-13) days. As an illustrative example, day-by-day sleep duration from 4 random participants is shown in Figure 1. Large variability can be seen in day-by-day data within each 14-day period before the PHQ-8 assessment. Table 3 illustrates how the ICC and Bland-Altman analysis result varied with the number of days for sleep duration using data from all the included participants. We can see that the absolute agreement improved when more days were included. Specifically, the ICC reached 0.9 with ≥8 days for the ICC of the mean as well as the ICC of the median. Table 4 shows the minimal number of days achieving an ICC >0.9 and the corresponding Bland-Altman analysis results for all the considered features. Most features showed high ICC when ≥8 days were included, except for maximum traveled distance from home, sleep efficiency, sleep offset time, unlock duration sum, and WASF. In general, the ICC of the median required more days to achieve absolute agreement than the ICC of the mean.

**Figure 1.** Sleep duration from 4 random participants. Weekend 0 and 1 (denoted by color only) are weekdays and weekends, respectively. The 8-item Patient Health Questionnaire depression (PHQ-8) scale order j=1, 2, 3, 4 (denoted by marker only) represents the j-th 14-day period (before the PHQ-8 assessment) having the same PHQ-8 scores. Mean and median were calculated within the same PHQ-8 score.

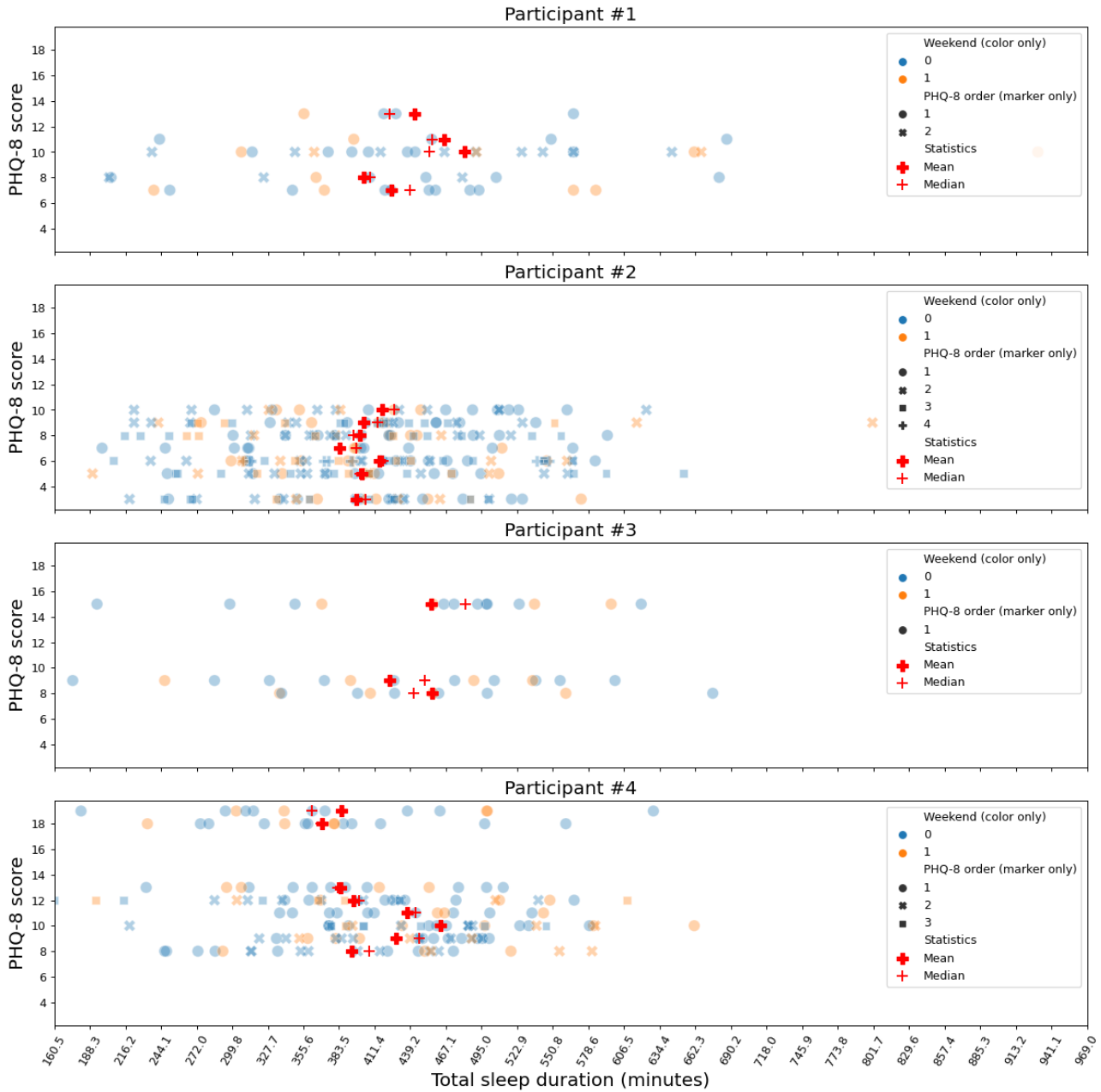

**Table 3.** The impact of the number of days (1-13) of data included on the intraclass correlation coefficient (ICC) and Bland-Altman analysis result when calculating 14-day summary statistics for sleep duration.

| Day | ICC mean | ICC median | Precision[a] (ICC mean) | Precision ratio (ICC mean) | Bias (ICC median) | Precision (ICC median) | Precision ratio (ICC median) |
| --- | --- | --- | --- | --- | --- | --- | --- |
| 1 | 0.32 | 0.42 | 151.87 | 0.34 | 2.03 | 155.09 | 0.34 |
| 2 | 0.55 | 0.64 | 103.18 | 0.23 | 2.03 | 107.85 | 0.24 |
| 3 | 0.67 | 0.70 | 80.66 | 0.18 | 0.99 | 94.17 | 0.21 |
| 4 | 0.75 | 0.80 | 66.60 | 0.14 | 0.99 | 73.66 | 0.16 |
| 5 | 0.80 | 0.82 | 56.51 | 0.12 | 0.59 | 68.80 | 0.15 |
| 6 | 0.85 | 0.87 | 48.64 | 0.10 | 0.59 | 56.04 | 0.12 |
| 7 | 0.88 | 0.88 | 42.12 | 0.09 | 0.35 | 53.17 | 0.11 |
| 8 | 0.91 | 0.92 | 36.48 | 0.08 | 0.35 | 43.63 | 0.09 |
| 9 | 0.93 | 0.93 | 31.39 | 0.07 | 0.20 | 40.95 | 0.09 |
| 10 | 0.95 | 0.95 | 26.64 | 0.05 | 0.20 | 32.87 | 0.07 |
| 11 | 0.96 | 0.96 | 21.99 | 0.04 | 0.09 | 29.74 | 0.06 |
| 12 | 0.97 | 0.98 | 17.19 | 0.03 | 0.09 | 21.81 | 0.04 |
| 13 | 0.98 | 0.98 | 11.68 | 0.02 | 0.00 | 16.86 | 0.03 |

[a]Bias (ICC mean) were all 0.

**Table 4.** The minimal number of days of data for different features when the intraclass correlation coefficient (ICC) of the mean and median reached 0.9[a].

|  | ICC mean >0.9 (d) | Precision[b] (ICC mean) | ICC median >0.9 (d) | Bias (ICC median) | Precision (ICC median) |
| --- | --- | --- | --- | --- | --- |
| Homestay duration (min) | 8 | 85.8 | 9 | −0.3 | 115.5 |
| Maximum traveled distance from home (km) | 12 | 34.0 | 10 | −0.2 | 13.8 |
| Step count | 6 | 2341.6 | 8 | −32.3 | 2201.0 |
| Sleep duration (min) | 8 | 36.5 | 8 | 0.4 | 43.7 |
| Time in bed (min) | 8 | 40.1 | 8 | 0.4 | 48.4 |
| Sleep efficiency | 9 | 0.01 | 10 | 0.0 | 0.01 |
| WASO[c] (min) | 8 | 8.2 | 8 | −0.2 | 10.7 |
| WASF[d] (min) | N/A[e] | N/A | 12 | 0.0 | 1.4 |
| Sleep onset time (h) | 6 | 0.8 | 6 | 0.0 | 0.9 |
| Sleep offset time (h) | 11 | 0.9 | 8 | 0.0 | 0.9 |
| Unlock times | 2 | 17.7 | 2 | −0.9 | 18.4 |
| Unlock duration sum (min) | 4 | 70.0 | 4 | −2.9 | 80.3 |
| Unlock duration min (min) | 11 | 0.9 | 12 | 0.0 | 0.6 |
| Unlock duration max (min) | 6 | 19.8 | 8 | −0.7 | 17.3 |
| Median interval between 2 unlocks (min) | 8 | 20.0 | 8 | −0.6 | 16.4 |
| Social app use times | 3 | 24.6 | 4 | −0.7 | 23.0 |
| Social app use duration (min) | 5 | 28.3 | 6 | −1.0 | 29.5 |
| Attention app use times | 5 | 6.9 | 6 | −0.3 | 6.7 |
| Attention app use duration (min) | 6 | 10.5 | 7 | −0.3 | 11.0 |
| Game app use times | 6 | 8.3 | 6 | −0.3 | 9.0 |
| Game app use duration (min) | 5 | 40.1 | 6 | −0.7 | 42.0 |

[a]The corresponding Bland-Altman analysis result is also presented.
[b]Bias for ICC mean were all 0.
[c]WASO: wakefulness after sleep onset.
[d]WASF: wakefulness after sleep offset.
[e]N/A: not applicable (as the required absolute agreement was not achieved).

## Cross-Sectional and Longitudinal Correlations Between the Extracted Features and Depression Symptoms

We included 14-day periods (before the PHQ-8 assessment) where a minimum of 8 days of data were found and used the mean of the features for all data streams. Table 5 shows the Spearman correlation coefficients using individual-mean data, repeated measures correlation coefficients, and the results from linear mixed effects models. The rankings in the cross-sectional and longitudinal correlations differed for some of the features; for instance, sleep onset time ranked first cross-sectionally but much lower longitudinally, whereas WASO ranked second longitudinally but much lower cross-sectionally. Nevertheless, features such as step count ranked high both longitudinally and cross-sectionally. In general, longitudinal correlation coefficients were smaller than the cross-sectional ones.

Table 5. The cross-sectional and longitudinal associations between the features and the 8-item Patient Health Questionnaire scores[a].

| Feature | Cross-sectional | | Longitudinal | | | | |
|---|---|---|---|---|---|---|---|
| | Individual-mean correlation coefficient | Ranking | Repeated measures correlation coefficient | Ranking | $P$ value[b] | $t$ value[b] | Ranking[b] |
| Sleep onset time | 0.24 | 1 | –0.05 | 10 | .28 | –1.08 | 20 |
| Step count | –0.19 | 2 | –0.14 | 1 | <.001 | –6.34 | 1 |
| WASF[c] | –0.17 | 3 | –0.01 | 21 | .23 | –1.2 | 17 |
| Maximum traveled distance from home | –0.14 | 4 | –0.03 | 16 | .06 | –1.85 | 12 |
| Homestay duration | 0.10 | 5 | 0.11 | 3 | <.001 | 5.09 | 3 |
| Unlock duration max | 0.08 | 6 | 0.05 | 12 | .02 | 2.38 | 9 |
| Attention times | –0.07 | 7 | 0.07 | 7 | .04 | 2.04 | 11 |
| Unlock times | –0.06 | 8 | 0.05 | 13 | .13 | 1.51 | 15 |
| Sleep offset time | 0.06 | 9 | –0.03 | 18 | .26 | –1.12 | 18 |
| Median interval between 2 unlocks | 0.05 | 10 | 0.02 | 19 | .28 | 1.09 | 19 |
| Attention duration | –0.03 | 11 | 0.05 | 11 | .07 | 1.84 | 13 |
| Social times | –0.03 | 12 | 0.05 | 14 | .09 | 1.71 | 14 |
| WASO[d] | 0.03 | 13 | 0.13 | 2 | <.001 | 5.79 | 2 |
| Sleep efficiency | 0.02 | 14 | –0.1 | 4 | <.001 | –3.87 | 4 |
| Time in bed | –0.02 | 15 | 0.09 | 5 | <.001 | 3.75 | 5 |
| Game times | –0.02 | 16 | –0.01 | 20 | .93 | 0.09 | 21 |
| Unlock duration sum | 0.02 | 17 | 0.06 | 9 | .01 | 2.58 | 8 |
| Social duration | 0.01 | 18 | 0.07 | 8 | <.001 | 2.93 | 7 |
| Sleep duration | –0.01 | 19 | 0.08 | 6 | <.001 | 3.21 | 6 |
| Unlock duration min | 0.01 | 20 | 0.03 | 17 | .14 | 1.46 | 16 |
| Game duration | –0.01 | 21 | 0.05 | 15 | .04 | 2.04 | 10 |

[a]The ranking in each type of associations is presented.
[b]Results from linear mixed effects models.
[c]WASF: wakefulness after sleep offset.
[d]WASO: wakefulness after sleep onset.

### Clustering of Participants Based on Behavioral Differences

The PCA-explained variance ratio is given in Figure 2A; the requirement of the variance was met when including 7 features, which led to 9 features being selected in the next step of the PFA. After running the PFA, the selected features were sleep duration, step count, WASF, sleep offset time, sleep efficiency, unlock duration sum, unlock duration min, attention times, and game times. Figure 2B shows silhouette coefficients varying with the number of clusters, where 3 clusters achieved the best clustering performance. The best silhouette coefficient score achieved was 0.17, suggesting some overlap among the clusters. The derived centers of the 3 clusters are presented in Table 6. We can see that sleep duration increased across all 3 clusters with the depression symptom severity. For step count, sleep efficiency, unlock duration sum, unlock duration min, and game times, participants in different clusters experienced changes in opposite directions with changes in symptom severity. Specifically, when experiencing depression, participants in cluster 1 slept longer, walked less, and woke up later; participants in cluster 2 changed behaviors marginally; and participants in cluster 3 reduced the time and frequency of using their smartphone. The weights for these 3 clusters were 0.06, 0.73, and 0.21, respectively. The mean and SD values of clustered data after standardization are presented in Figure 3. After visual inspection, cluster 1 showed a fair separability from clusters 2 and 3. We did not find group difference in age and length of education among the 3 clusters. The repeated measures correlation coefficients within each cluster are given in Table 7. Differences in the strengths of repeated measures correlation coefficients among the 3 clusters can be found in features such as step count, which is in line with the findings in Table 6; yet, for some of the other features, the correlation coefficients revealed different signs (directions of change) compared with the center of that cluster. The repeated measures correlation coefficients in each cluster also differed from those derived without clustering ([Table 5](#)).

**Figure 2.** (A) Explained variance ratio in the principal component analysis on the processed behavioral difference (quantified by the features) between periods of high and low 8-item Patient Health Questionnaire depression scale scores. (B) Silhouette coefficients in the clustering analysis using the Gaussian mixture model.

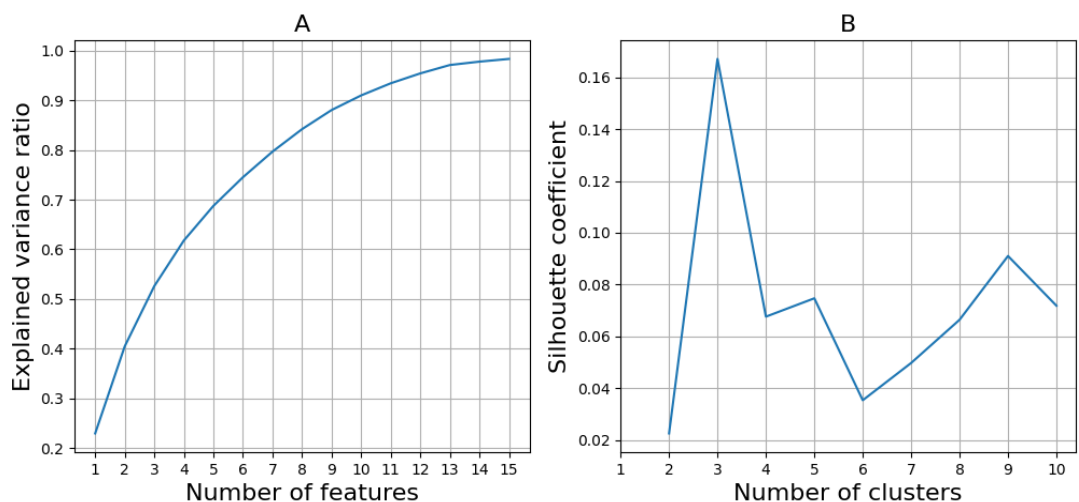

**Table 6.** The centers of the identified 3 clusters from the Gaussian mixture model.

|  | Cluster 1 | Cluster 2 | Cluster 3 |
| --- | --- | --- | --- |
| Sleep duration | 0.27 | 0.03 | 0.05 |
| Step count | −0.26 | −0.06 | 0.04 |
| WASF[a] | −0.05 | −0.02 | 0.01 |
| Sleep offset time | 0.26 | 0.04 | −0.07 |
| Sleep efficiency | 0.11 | −0.02 | −0.04 |
| Unlock duration sum | −0.05 | 0.02 | −0.14 |
| Unlock duration min | 0.06 | 0.02 | −0.19 |
| Attention times | 0.04 | 0.02 | −0.07 |
| Game times | 0.16 | 0.02 | −0.10 |

[a]WASF: wakefulness after sleep offset.

**Figure 3.** Parallel plot for the mean and SD of the identified 3 clusters after standardization. WASF: wakefulness after sleep offset.

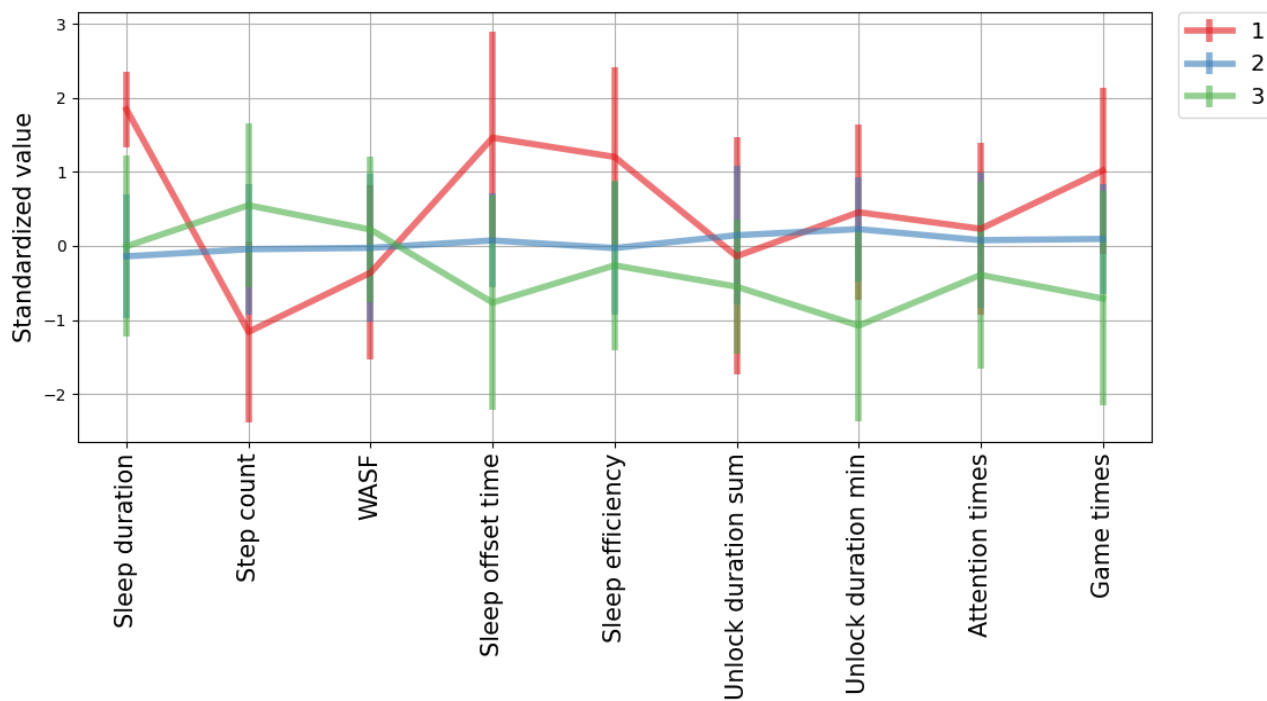

**Table 7.** The repeated measures correlation coefficients between the features and the 8-item Patient Health Questionnaire scores in each cluster[a].

|  | Cluster 1 | Cluster 2 | Cluster 3 |
| --- | --- | --- | --- |
| Sleep duration | 0.05 | 0.08 | 0.14 |
| Step count | −0.36 | −0.16 | −0.14 |
| WASF[b] | −0.12 | −0.05 | 0.04 |
| Sleep offset time | −0.11 | 0.02 | −0.04 |
| Sleep efficiency | −0.13 | −0.10 | −0.04 |
| Unlock duration sum | 0.00 | 0.07 | 0.21 |
| Unlock duration min | −0.06 | −0.03 | 0.34 |
| Attention times | 0.15 | 0.09 | 0.09 |
| Game times | 0.26 | −0.07 | 0.10 |

[a]The features were selected using principal feature analysis.
[b]WASF: wakefulness after sleep offset.

## Discussion

### Principal Findings

In this work, we presented an exploration of 3 challenges when analyzing digital phenotypes to associate with, and predict, depression symptom severity. These included the impact of data missingness on feature quality, the nature of the correlation between symptom severity and digital phenotypes, and the heterogeneity with which depression manifests itself. We examined the impact of data missingness when calculating summary statistics for mHealth-derived features over a 14-day window and found that 8 days of data were necessary for most of the features. We also demonstrated that the cross-sectional and longitudinal correlations between features and depression symptom severity as measured by the PHQ-8 differed considerably for some of the features. Furthermore, we identified the heterogeneity with which depression manifests itself in our cohort, showing that participants could be stratified into 3 clusters based on the changes in mHealth-derived features between periods of high and low PHQ-8 scores.

Missing data are inevitable in remote monitoring and mHealth studies for reasons that include device-wearing habits, device malfunction, and network connectivity. An inadequate number of days when calculating the summary statistics will introduce unexplained variability, complicating data analysis [52]. Existing studies have used an arbitrary threshold (eg, 7 or 12 days) [12,20] to determine data usability or have not reported a threshold, which might lead to discrepancies in findings. To the best of our knowledge, no studies have quantitatively and comprehensively investigated the impact of data missingness when using remote monitoring technology in assessing depression symptom severity. Our study filled this gap by demonstrating that a wide range of 2 to 12 feature-specific random days out of 14 were necessary, whereas ≤8 random days were adequate for most of the features. This finding can promote the reproducibility of data analysis and provide flexibility in data collection. Features with large day-to-day variability within participants often needed more data to be representative than features that were relatively stable across days. The reason why the ICC of the median required more days to achieve absolute agreement than the ICC of the mean might be related to the fact that random errors were suppressed when calculating the mean. However, when the day-to-day variability was pronounced, as in the case of maximum traveled distance from home, the ICC of the median yielded better absolute agreement. It is interesting to observe that features extracted from the same data stream can differ in the minimal days required, such as homestay duration and maximum traveled distance from home. It should be noted that although data missingness poses challenges to feature quality and data analysis, it might suggest relevant information about depressive status [53].

When analyzing data collected from observational studies with repeated measurements, it is necessary to distinguish between cross-sectional and longitudinal correlations, especially considering the case of the Simpson paradox, where contradictory patterns are found in higher and lower levels of analysis [47]. The findings relating to the difference between the longitudinal and cross-sectional correlations suggest that different feature subsets are needed for population screening and patient continuous monitoring purposes. Accordingly, different models may need to be built for depression symptom severity assessments, depending on the specific purposes. The reason why features behaved differently in the cross-sectional and longitudinal associations might be related to demographic factors such as age, gender, and occupation [54]; for instance, sleep patterns have been reported to differ by age and gender [54]. Future work may investigate the nature and strength of correlations in different demographic subgroups in larger data sets to further understand the utility of the features. The rankings from the repeated measures correlation and linear mixed effects model can be different, which may be related to the way variance is handled. The repeated measures correlation analyzes intraindividual variance, whereas the linear mixed effects model can simultaneously analyze both intra- and interindividual variance using partial pooling [47]. Features such as step count were sensitive in detecting changes in depression symptom severity and in screening people with depression in line with previous studies [55,56]. This might be associated with inverse bidirectional relationships between depression and physical activity [57,58]. Furthermore, a potential protective relationship between physical activity and depression might also be a contributing factor [59]. However, overall, the cross-sectional and longitudinal correlations were generally small. This suggests that the value of these representative features for predicting depression depends very much on the specific application, and more sophisticated features and other sensor modalities are needed.

The discovery of the 3 clusters suggests that there was heterogeneity in people with depression in terms of behavioral patterns. This finding might explain why correlations between the PHQ-8 scores and features were relatively small and why varied predictive performances have been reported in other studies [15,60]. In addition, the individual heterogeneity also suggests that personalized predictive models for remote depression assessment are needed. Clustering individuals before building predictive models, coupled with selecting features that are better at capturing longitudinal correlations, might improve model performance. In this study, the clustering centers showed that smartphone and app use could change in opposite directions for different individuals when depressed. One speculation, based on the items in the PHQ-8, is that some people lost interest in doing things, including using their smartphones and apps, whereas others resorted to using their smartphones more when having difficulty focusing on other activities. When considered together, changes in different features in the same cluster might also shed light on behavioral changes; for example, participants in cluster 1 had longer sleep duration and less physical exercise when becoming more depressed. The noticeable overlap between the clusters and high variance might be due to factors such as sample size. It is of particular interest to note that some of the participants experienced marginal changes in their behavioral patterns when more depressed; additional relevant features and sensors might also help to improve the understanding of heterogeneity of behavioral patterns in people with MDD. It should be noted that the discovered clusters had some overlap, as suggested by the derived silhouette coefficients; future work may investigate whether including additional complementary features in larger data sets could help to improve cluster cohesion. Repeated measures correlation coefficients within each cluster were also consistent with some of the findings from clustering analysis using the Gaussian mixture model. The reason for the discrepancies between the sign of the correlation and that of the center of the cluster might be 2-fold. First, when the correlation coefficients were small, the direction of change might be subject to the way data were analyzed. Furthermore, the relationship between the features and PHQ-8 scores might be more complicated than being monotonic, which can be captured by the correlation analysis. Future work might examine further the associations in different PHQ-8 score ranges. The difference in the repeated measures correlation coefficients in each cluster and those derived without clustering further highlights the heterogeneity in the depression symptoms. Overall, the discovery of the heterogeneity might suggest that cluster-specific predictive models are needed when using mHealth-derived features.

There have been works exploring the RADAR-MDD cohort from different perspectives and in different ways. We previously elaborated issues in relation to recruitment, retention, data availability, and predictors of engagement [30]. We also extracted features derived from different data streams, including Fitbit sleep, smartphone Bluetooth, GPS, and acceleration, and associated them with depression symptom severity [16,17,20,34,61]. These works examining associations often used heuristically defined thresholds to determine data and feature quality, and the derived correlations and associations were focused on either the longitudinal or cross-sectional aspect.

Furthermore, the heterogeneity with which depression manifests itself was not considered. This work adds to these existing works analyzing the same cohort by investigating these factors that may complicate the relationship between digital phenotypes and depression symptom severity. The findings in this work may also help to interpret findings in these existing works using the same cohort.

## Limitations

There are some limitations to this study. First, the sleep stages provided by the Fitbit devices may differ from the gold standard using polysomnography [62]; for example, inactive status might be misclassified as sleep, which leads to overestimation of sleep duration [63]. Nevertheless, wearable devices provide a means to unobtrusively monitor individuals for much extended time periods with comfort and to capture real-world data in a home setting with greater ecological validity. Future work may use devices that provide raw signals to have better control of algorithmic details, facilitating deeper insights. Second, the definitions of periods of depression (high depression scores) and no depression (low depression scores) were based on the PHQ-8 with a cutoff of 10, which potentially led to inaccurate classifications. When scores were close to 10, they were classified into no-depression and depression periods, but the closeness in scores was overlooked. However, the focus of this study was to quantify the difference in the behavioral patterns in which a threshold was needed. Future work may investigate the impact of scores or use other criteria to determine depression. Other criteria (eg, the Center for Epidemiologic Studies Depression Scale) may also provide deeper insights into how individual depression symptoms manifest into subsets of the entire spectrum of MDD symptoms [64]. Third, participants in this study were requested to maintain their regular routines, which included medication use. The use of medication could be a confounding factor that affected their behavioral patterns both in this study and in real-life situations. In the future, larger studies that frequently gather medication information may further investigate the role of medication use in a person's behavioral patterns when experiencing symptoms of depression.

## Conclusions

This study investigated 3 considerations when using digital phenotypes to test for associations with, and predict, depression symptom severity. We examined the impact of the number of days where data were available for calculating representative summary statistics and found that the minimal number of 8 (range 2-12) days from a 14-day window was sufficient for most of the features. The strength of associations varied according to whether features were being considered in a cross-sectional or longitudinal context. Furthermore, 3 clusters were identified when analyzing behavioral patterns using mHealth-derived features, which demonstrated the heterogeneity with which depression manifests itself. This work supports the need to consider stratified analyses in future work. Overall, the analysis presented here highlights important issues to be considered when analyzing data collected from wearable devices and smartphones to assess depression symptom severity to inform future studies.

## Acknowledgments

This study represents independent research partly funded by the National Institute for Health and Care Research (NIHR) Maudsley Biomedical Research Centre (BRC) at South London, Maudsley National Health Service (NHS) Foundation Trust, King's College London, and European Union/European Federation of Pharmaceutical Industries and Associations (EFPIA) Innovative Medicines Initiative (IMI)-2 Joint Undertaking (Remote Assessment of Disease and Relapse–Central Nervous System [RADAR-CNS]: 115902). This communication reflects the views of the RADAR-CNS consortium, and neither IMI nor the European Union and EFPIA are liable for any use that may be made of the information contained herein. Participant recruitment in Amsterdam, the Netherlands, was partially accomplished through a Dutch web-based registry that facilitates participant recruitment for neuroscience studies [65]. It is funded by ZonMw-Memorabel (73305095003), a project in the context of the Dutch Deltaplan Dementie, Gieskes-Strijbis Foundation, the Alzheimer's Society in the Netherlands (Alzheimer Nederland), and Brain Foundation Netherlands (Hersenstichting). This study has also received support from Health Data Research UK (funded by the UK Medical Research Council [MRC]), Engineering and Physical Sciences Research Council, Economic and Social Research Council, Department of Health and Social Care (England), Chief Scientist Office of the Scottish Government Health and Social Care Directorates, Health and Social Care Research and Development Division (Welsh Government), Public Health Agency (Northern Ireland), British Heart Foundation, Wellcome Trust, and the NIHR University College London Hospitals Biomedical Research Centre.

The RADAR-CNS project has received funding from the IMI-2 Joint Undertaking (115902). This Joint Undertaking receives support from the European Union's Horizon 2020 Research and Innovation Program and EFPIA. The funding bodies have not been involved in the design of the study, the collection or analysis of data, or the interpretation of data. The views expressed are those of the author or authors and not necessarily those of the NHS, the NIHR, or the Department of Health and Social Care. The authors thank all members of the RADAR-CNS patient advisory board for their contribution to the device selection procedures and their invaluable advice throughout the study protocol design. This research was reviewed by a team with experience of mental health problems and their carers, who have been specially trained to advise on research proposals and documentation through the Feasibility and Acceptability Support Team for Researchers (FAST-R), a free confidential service in England provided by the NIHR Maudsley BRC via King's College London and the South London and Maudsley NHS Foundation Trust. The authors thank all Genetic Links to Anxiety and Depression Study volunteers for their participation and gratefully acknowledge the NIHR BioResource, NIHR BioResource Centres, NHS Trusts, and staff for their contribution. The authors also acknowledge NIHR BRC, King's College London, the South London and Maudsley NHS Trust, and King's Health Partners. The authors thank the NIHR, NHS Blood and Transplant, and Health Data Research UK as part of the Digital Innovation Hub Program.

CO is supported by the UK MRC (MR/N013700/1) and King's College London member of the MRC Doctoral Training Partnership in Biomedical Sciences. RJBD is supported by the following: (1) NIHR BRC at the South London and Maudsley NHS Foundation Trust and King's College London; (2) Health Data Research UK, which is funded by the UK MRC, Engineering and Physical Sciences Research Council, Economic and Social Research Council, Department of Health and Social Care (England), Chief Scientist Office of the Scottish Government Health and Social Care Directorates, Health and Social Care Research and Development Division (Welsh Government), Public Health Agency (Northern Ireland), British Heart Foundation, and Wellcome Trust; (3) the BigData@Heart consortium, funded by the IMI-2 Joint Undertaking (116074); this Joint Undertaking receives support from the European Union's Horizon 2020 Research and Innovation Program and EFPIA, and it is chaired by DE Grobbee and SD Anker, partnering with 20 academic and industry partners and the European Society of Cardiology; (4) the NIHR University College London Hospitals Biomedical Research Centre; (5) the NIHR BRC at the South London and Maudsley NHS Foundation Trust and King's College London; (6) the UK Research and Innovation London Medical Imaging and Artificial Intelligence Centre for Value Based Healthcare; and (7) the NIHR Applied Research Collaboration (ARC) South London at King's College Hospital NHS Foundation Trust.

## Data Availability

The data sets used for this study can be made available through reasonable requests to the Remote Assessment of Disease and Relapse–Central Nervous System (RADAR-CNS) consortium. Please email the corresponding author for details.

## Conflicts of Interest

SV and VAN are employees of Janssen Research and Development LLC. PA is employed by the pharmaceutical company H Lundbeck A/S. DCM has accepted honoraria and consulting fees from Otsuka Pharmaceuticals Co, Ltd; Optum Behavioral Health; Centerstone Research Institute; and the One Mind Foundation; has received royalties from Oxford Press; and has an ownership interest in Adaptive Health, Inc. MH is the principal investigator of Remote Assessment of Disease and Relapse–Central Nervous System (RADAR-CNS), a private-public precompetitive consortium that receives funding from Janssen Research and Development LLC, UCB, H Lundbeck A/S, MSD, and Biogen. All other authors declare no other conflicts of interest.

## Abbreviations

**CIBER:** Centro de Investigación Biomédica en Red
**ICC:** intraclass correlation coefficient
**IMI:** Innovative Medicines Initiative
**MDD:** major depressive disorder
**mHealth:** mobile health
**NHS:** National Health Service
**PCA:** principal component analysis
**PFA:** principal feature analysis
**PHQ-8:** 8-item Patient Health Questionnaire
**RADAR-base:** Remote Assessment of Disease and Relapse–base
**RADAR-CNS:** Remote Assessment of Disease and Relapse–Central Nervous System
**RADAR-MDD:** Remote Assessment of Disease and Relapse–Major Depressive Disorder
**REM:** rapid eye movement
**VUMC:** Vrije Universiteit Medisch Centrum
**WASF:** wakefulness after sleep offset
**WASO:** wakefulness after sleep onset